\documentclass[10pt,a4paper]{article}

\newcommand{\xa}{\mathbf{x}^a}
\newcommand{\xb}{\mathbf{x}^b}
\newcommand{\xt}{\mathbf{x}^t}
\newcommand{\xx}{\mathbf{x}}
\newcommand{\yo}{\mathbf{y}^o}

\newcommand{\xaref}{\xa_{ref}}

\newcommand{\oA}{\mathbf{A}}
\newcommand{\oB}{\mathbf{B}}
\newcommand{\oC}{\mathbf{C}}

\newcommand{\oH}{\mathbf{H}}
\newcommand{\oI}{\mathbf{I}}
\newcommand{\oK}{\mathbf{K}}

\newcommand{\oR}{\mathbf{R}}

\usepackage{graphicx}% Include figure files
\usepackage{dcolumn}% Align table columns on decimal point
\usepackage{bm}% bold math
\begin{document}

\title{Differential influence of instruments in nuclear core activity evaluation
by data assimilation}

\author{Bertrand Bouriquet $^1$ \footnote{bertrand.bouriquet@cerfacs.fr}
   \and Jean-Philippe Argaud $^{2,1}$
   \and Patrick Erhard $^2$
   \and S\'ebastien Massart $^1$
   \and Ang\'elique Pon{\c c}ot $^2$
   \and Sophie Ricci $^1$
   \and Olivier Thual $^{1,3}$}

\maketitle

\footnotetext[1]{
Sciences de l'Univers au CERFACS, URA CERFACS/CNRS No~1875,
42 avenue Gaspard Coriolis,
F-31057 Toulouse Cedex 01 - France
}
\footnotetext[2]{
Electricit\'e de France,
1 avenue du G\'en\'eral de Gaulle,
F-92141 Clamart Cedex - France
}
\footnotetext[3]{
Universit\'e de Toulouse, INPT, UPS, IMFT,
All\'ee Camille Soula,
F-31400 Toulouse - France
}

\begin{abstract}

The global activity fields of a nuclear core can be reconstructed using data
assimilation. Data assimilation allows to combine measurements from instruments,
and information from a model, to evaluate the best possible activity within the
core. We present and apply a specific procedure which evaluates this influence
by adding or removing instruments in a given measurement network (possibly
empty). The study of various network configurations of instruments in the
nuclear core establishes that influence of the instruments depends both on the
independant instrumentation location and on the chosen network. 

{\bf Keywords:}
Data assimilation, neutronic, activities reconstruction,  nuclear in-core
measurements, data acquisition network
 
\end{abstract}

%-------------------------------------------------------------------------------
\section{Introduction}

Data assimilation has been widely developed in earth sciences and especially
meteorology. Introducing data assimilation was an important step to improve
weather forecasts \cite{Parrish92,Todling94,Ide97}. Such a technique is now used
operationally in all meteorological office. The efficiency of data assimilation
for field reconstruction has already been demonstrated in several articles in
meteorology \cite{era40,Kalnay96,Huffman97}. 

The purpose is to use this technique to make an optimal reconstruction of the
activity field in a nuclear core using both measurements and information coming
from a numerical model. This approach was also used  for nuclear core  neutronic
state evaluation, and particularly nuclear activity field
\cite{Massart07,Bouriquet2010}.

In \cite{Bouriquet2010}, the authors demonstrate how this method is tolerant to
instrument loss and that this effect is related to the instrument repartition. 
The aim of the present article is to go further in the understanding of the
instruments location and accuracy effect, and focus more specifically on the
individual contribution of each instrument. To achieve this goal, the method is
based on the statistical occurrence of instruments with respect to the global
quality of the reconstruction associated. 

The next section describes the general data assimilation concepts that are used
all along this article. Then the parametrisation of data assimilation components
are presented in details. From those bases of data assimilation, the specific
methodology used to track the individual influence of an instrument on data
assimilation is described. Finally, using all previous information, an
investigation is done on  the instrument location and error modelling effect on
the influence of instrument in the core is done.

%-------------------------------------------------------------------------------

\section{Data assimilation\label{sec:da}}

Here are briefly introduced the useful data assimilation key points to understand
their use as applied here \cite{Talagrand97,Kalnay03,Bouttier99}. But data
assimilation is a wider domain and these techniques are for example the keys of
the nowadays meteorological operational forecast \cite{Rabier2000}. This is
through advanced data assimilation methods that the weather forecast has been
drastically improved during the last 30 years. Those techniques use all the
available data, such as satellite measurements, as well as sophisticated
numerical models. 

The ultimate goal of data assimilation methods is to estimate the inaccessible
true value of the system state,  $\xt$ where the $t$ index stands for "true
state" in the so colled "control space". The basic idea of most of data
assimilation method is to combine information  from an \textit{a priori} on the
state of the system (usually called $\xb$, with $b$ for "background"), and 
measurements (referenced as $\yo$). The background is usually the result of
numerical simulations but can also be derived from any {\it a priori} knowledge.
The result of data assimilation is called the analysis, denoted by $\xa$, and it
is an estimation of the true state $\xt$ researched.   

The control and observation spaces are not necessary the same, a bridge between
them needs to be build. This is the  observation operator $H$ that transform
values from the space of the background to the space of observations. For our
data assimilation purpose we will use its linearisation $\oH$ around the
observation values. The inverse operation to go from space of observations to
space of the background is given by the transpose $\oH^T$ of $\oH$.

Two other ingredients are necessary. The first one is the covariance matrix of
observation errors, defined as $\oR=E[(\yo-H(\xt)).(\yo-H(\xt))^T]$ where $E[.]$
is the mathematical expectation. It can be obtained from the known errors on
unbiased measurements which means $E[\yo-H(\xt)]=0$. The second one is the
covariance matrix of background errors, defined as
$\oB=E[(\xb-\xt).(\xb-\xt)^T]$. It represents the error on the {\it a priori}
state, assuming it to be unbiaised following the $E[\xb-\xt]=0$ no biais
property. There are many ways to get this {\it a priori} state and background
error matrices. However, those matrix are commonly the output of a model and an
evaluation of accuracy, or the result of expert knowledge. 

It can be proved, within this formalism, that the Best Unbiased Linear Estimator
(BLUE) $\xa$, under the linear and static assumptions, is given by the following
equation:
\begin{equation}\label{xa}
\xa = \xb + \oK \big(\yo- H\xb\big),
\end{equation}
where $\oK$ is the gain matrix:
\begin{equation} \label{Kmat}
\oK = \oB\oH^T (\oH\oB\oH^T + \oR)^{-1}.
\end{equation}
Moreover, we can get the analysis error covariance matrix $\oA$, characterising
the analysis errors $\xa-\xt$. This matrix can be expressed from $\oK$ as:
\begin{equation}
\oA = (\oI - \oK\oH)\oB,
\end{equation}
where $\oI$ is the identity matrix.

It is worth noting that solving Equation \ref{xa} is, if the probability
distribution is Gaussian, equivalent to minimise the following function
$J(\xx)$, $\xa$ being the optimal solution:
\begin{equation}\label{J}
J(\xx) = (\xx-\xb)^T \oB^{-1}(\xx-\xb) \medskip + \big(\yo-\oH\xx\big)^T \oR^{-1} \big(\yo-\oH\xx\big).
\end{equation}

This minimisation is known in data assimilation as 3D-Var methodology
\cite{Talagrand97}.

%-------------------------------------------------------------------------------

\section{Data assimilation implementation}

The framework of this study is the standard configuration of a 900 MWe nuclear
Pressurized Water Reactor (PWR900). To perform data assimilation, both
simulation code and data are needed. For the simulation code, the EDF
experimental calculation code for nuclear core COCAGNE in a standard
configuration is used. The description of the basic features of this model are
done in Section \ref{model}.

To have a good understanding of the instrumentation effect on nuclear activity
reconstruction, various kind of configurations  are studied, even some that do
not exist operationally and so cannot be tested experimentally. For that
purpose, synthetic data  are used, that allows to have an homogeneous approach
all along the document. Synthetic data is generated from a model simulation,
filtered thought an instrument model and noised according to a predefined
measurement error density function (Gaussian type). 

\begin{figure}[!ht]
\begin{center}
  \includegraphics[width=1.0\textwidth]{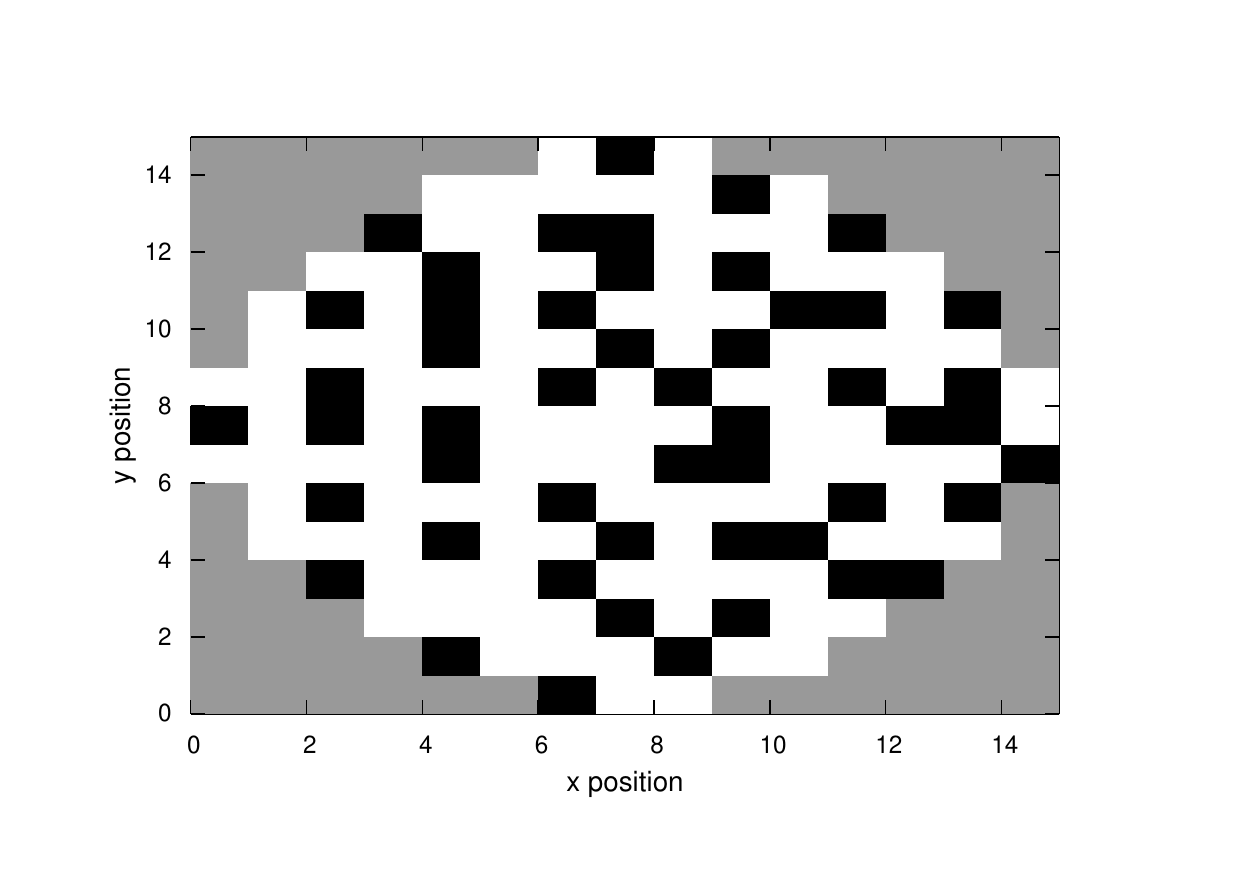}

  \caption{ The positions of MFC instruments in the nuclear core are localised
  in assemblies in black within the horizontal slice of the core. The assemblies
  without instrument are marked in white and the reflector, out of the reactive
  core, is in gray. \label{fig:figMFC}}

\end{center}
\end{figure} 

In the present case, we study the activity field reconstruction. An horizontal
slice of a PWR900 core is represented on the Figure~\ref{fig:figMFC}. There is a
total of $157$ assemblies within this core. Among those assemblies, $50$ are
instrumented with Mobiles Fissions Chambers (MFC). Those assemblies are divided
verticaly in $29$ vertical levels. Thus, the size of the control $\xx$ to be
estimated is $4553$ $(157 \times 29)$. The size of the observation vector $\yo$
is $1450$ $(50 \times 29)$. 

\subsection{Brief description of the nuclear core model\label{model}}

The aim of a neutronic code like COCAGNE is to evaluate the neutronic activity
field and all associated values within the nuclear core. This field depend on
the position in the core and on the neutron energy. To do such an evaluation,
the population of neutrons are divided in several groups of energy. In the
present case only two groups are taken into account giving the neutronic flux
$\Phi=(\Phi_1,\Phi_2)$ (even if the present code have no limit for the group
number). The material properties depend on the position in the core, as the
neutronic flux $\Phi$, identified by solving two-group diffusion equations
described by:
\begin{equation}\label{eq:cirep:eqn}
\left\{
\begin{array}{l}
\displaystyle - \mbox{div}(D_1\mbox{\textbf{grad}}\Phi_1) + (\Sigma_{a1} + \Sigma_r ) \Phi_1 = 
\frac{1}{k} \Big(\nu_1 \Sigma_{f1} \Phi_1 + \nu_2 \Sigma_{f2} \Phi_2 \Big) \\
\displaystyle -\mbox{div}(D_2\mbox{\textbf{grad}}\Phi_2) + \Sigma_{a2} \Phi_2 - \Sigma_r \Phi_1 = 0, \\
\end{array}
\right.
\end{equation}
where $k$ is the effective neutron multiplication factor, all the quantities and
the derivatives (except $k$) depend on the position in the core, $1$ and $2$ are
the group indexes, $\Sigma_r$ is the scattering cross section from group $1$ to
group $2$, and for each group, $\Phi$ is the neutron flux, $\Sigma_a$ is the
absorption cross section, $D$ is the diffusion coefficient, $\nu\Sigma_f$ is the
corrected fission cross section.

The cross sections also depend implicitly on the concentration of boron, which
is a substance added in the water used for the primary circuit to control the
neutronic fission reaction, throught a feedback supplementary model. This model
takes into account the temperature of the materials and of the neutron
moderator, given by external thermal and thermo-hydraulic models. A detailed
description of the core physic and numerical solving can be found in reference
\cite{Duderstatd76}.

The overall numerical resolution consists in searching for boron concentration
such that the eigenvalue $k$ is equal to $1$, which means that the nuclear power
production is stable and self-sustaining. It is named critical boron
concentration computation.

The activity in the core is obtained through a combination of the fluxes
$\Phi=(\Phi_1,\Phi_2)$, given on the chosen mesh of the core. Using homogeneous
materials for each assembly (for example $157$ in a PWR900 reactor), and
choosing a vertical mesh compatible with the core (usually $29$ vertical
levels), this result in a field of activity of size $157 \times 29=4553$ that
cover all the core.

\subsection{The observation operator $H$}

The $H$ observation operator is the composition of a selection and of a
normalisation procedure. The selection procedure extracts the values
corresponding to effective measurement among the values of the model space. The
normalisation procedure is a scaling of the value with respect to the geometry
and power of the core. The overall operation is non linear. However, with a
range of value compatible with assimilation procedure, we can calculate the
linear associated operator $\oH$. This observation matrix is a  $(4553 \times
1450)$ matrix.

\subsection{The background error covariance matrix $\oB$}

The $\oB$ matrix represents the covariance between the spatial errors for the
the background. The $\oB$ matrix is estimated as the product of a correlation
matrix $\oC$ by a scaling factor to set variances.  

The correlation $\oC$ matrix is built using a positive function that defines the
correlations between instruments with respect to a pseudo-distance in model
space. Positive functions have the property (via Bochner theorem) to build
symmetric defined positive matrix when they are used as matrix generator
\cite{Matheron70,Marcotte08}. In the present case, Second Order Auto-Regressive
(SOAR) function is used to prescribe the $\oC$ matrix. In such a function, the
amount of correlation depends from the euclidean distance between spatial
points. The radial and vertical correlation length ($L_r$ and $L_z$
respectively, associated to the radial $r$ coordinate and the vertical $z$
coordinate) have different values, which means we are dealing with a global
pseudo euclidean distance. The used function can be expressed as follow:
\begin{equation}  
C(r,z) = \left(1+\frac{r}{L_r}\right) \left(1+\frac{|z|}{L_z}\right)
         \exp{\left(-\frac{r}{L_r}-\frac{|z|}{L_z}\right)}.
\label{eqB}
\end{equation}
The matrix obtained by the above Equation \ref{eqB} is a correlation matrix. It
is then multiplied by a suitable variance coefficient to get covariance matrix.
This coefficient is obtained by statistical study of difference between model
and measurements in real case. In real cases, this value is set around a few
percent. In our case, the size of the $\oB$ matrix is related to the size of
model space so it is $(4553 \times 4553)$.

\subsection{The observation error covariance matrix $\oR$}

The observation error covariance matrix $\oR$ is approximated by a diagonal
matrix. It means we assume here that no significant correlation exists between
the measurement errors of the MFC. The usual modelling consists on taking the
diagonal values as a percentage of the observation values. This can be expressed
as:
\begin{equation}  
\oR_{jj} = \left( \alpha \yo_j \right)^2, \quad \forall j
\label{eqR}
\end{equation}
In our case, the $\alpha$ parameter is fixed according to the accuracy of the
measurement and the representative error associated to the instrument, and is
the same for all the diagonal coefficients.

This hypothesis of error dependence of the measure amplitude is an usual one.
Such an modelling error means that location with a large signal are the one with
the higher error. In the following, a constant observation variance is also
used. This leads to define the $\oR$ matrix diagonal according to the following
equation:
\begin{equation}  
\oR_{jj} = \beta^2, \quad \forall j
\label{eqRflat}
\end{equation}
The size of the $\oR$ matrix is related to the size of observation space, so it
is $(1450 \times 1450)$.

\section{Method of instrument quality attribution\label{method}}

To evaluate the quality of individual instruments with respect to the estimation
of the  core activity field, a reference state is needed. This state, denoted by
$\xaref$, corresponds to the data assimilation analysis using all the available
instruments. If $\xa$ denotes an analysis experiment performed with less
instruments than the maximum available ones, we denote by
$\epsilon=||\xa-\xaref||$ the norm of its difference with the reference
analysis. As $\xaref$ is the best available analysis, we expect that the norm
of the difference $\epsilon$ to increase as the number of instrument decreases.
The norm $\epsilon$ is thus a measure of the quality of a given experiment. 

We consider a set $P$ of $p$ experiments, for instance all those for which a
given number of instruments are removed. If $n$ is the maximum number of
available instruments and $k$ the fixed number of removed instruments, there are
$p=C_n^k$  experiments in  $P$. We then denote by $\epsilon_i$, for $i=1,...,p$,
the norms of the differences for each experiment with the reference analysis. We
can thus plot the histogram of the $\epsilon_i$ series. For instance, the
histograms corresponding to different choices of $k$ can be compared. The
average of the $\epsilon_i$ series increases when $p$ increases, since the
quality of the analyses decreases when instruments are removed. Its root mean
square provides information on the homogeneity of the instrument quality for the
field reconstruction. 

Dealing with the $\epsilon_i$ series can provide information about individual
instrument. If we select the $\epsilon_i$ values above a fixed threshold, for
instance the $10\%$ highest values, we can point out, for each corresponding
experiment $i$, the set of instruments which has not been removed. For each
instrument, we can count how many times it appears in such an experiment and
compare this number to the ones obtained with the other instruments. We can
thus assess that an instrument appearing a great number of times in experiments
leading to these high values of $\epsilon_i$ is of poor quality compared to an
instrument which seldom appears for these low quality experiments. 

\section{Application on  various core geometry}

\begin{figure}[!ht] 
\begin{center} 
  \includegraphics[width=\textwidth]{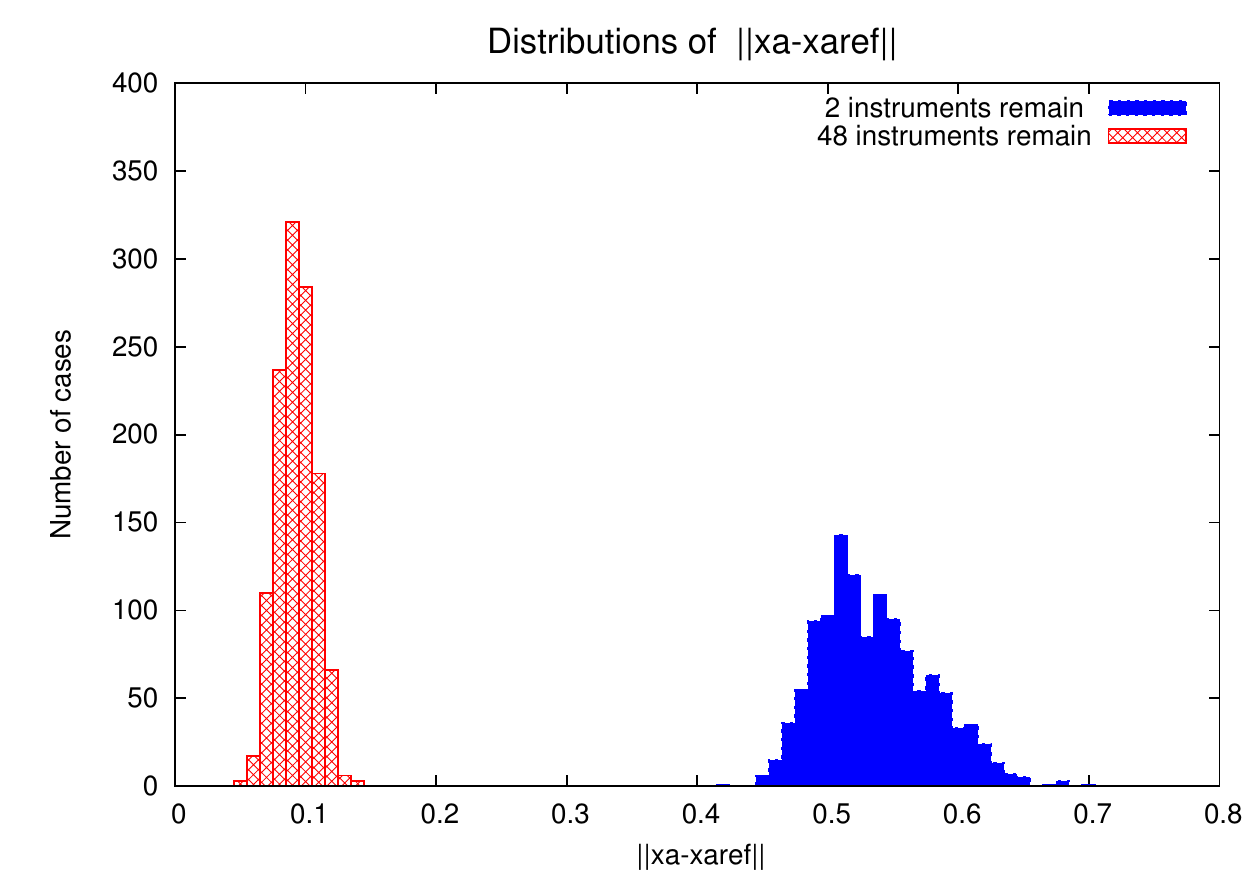} 

  \caption{Norm distribution for all the cases where only two instruments are
  suppressed or remain in the instrument network on a REP900 reactor
  \label{fig:inf-fig1}}

\end{center}
\end{figure} 

First of all, the behaviour of  $\epsilon_i$  for two extreme cases is examined:
when $2$ instruments are removed ({\it i.e.} $48$ remain) and when only 2
instruments remain. In both cases, $P$ sets contain  $p=1225$  case. This
corresponds to all possible combinations to  $C^{50}_{2}=C^{50}_{48}=1225$. On
Figure~\ref{fig:inf-fig1} are presented the distributions of the  
$\epsilon_i$'s for the cases where two instruments are lost ({\it i.e} $48$
remain) or only $2$ remain.

On Figure~\ref{fig:inf-fig1}, large differences between the two distributions
can be noticed. The shifting of the mean value between those two extreme cases
is logical as available information is dramatically changing. Thus the
difference of  $\epsilon_i$  between the two case is changing a lot on the mean.
However, the shape of the distribution is also changing. The distribution is a
very sharp one when $48$ instruments remain, whereas when only $2$ instruments
remain the distribution is broader.  The broadening of the distribution shows
that the instruments do not have the same influence on the activity field
reconstruction. If all instruments have had the same effect on data
assimilation, they would have been equivalent. Suppressing whatever instrument
would then change the mean value of the distribution of the norm but not the
shape. Ideally, this distribution shape should be a very sharp peak. In the
present case the distribution when $48$ instruments remain could be a good
candidate. Thus if all the instruments where equivalent, only a translation
between the distributions of $48$ instruments remains and $2$ instruments
remains would have been observed. However, transformation between the two
distributions proves that the instruments are not all equivalent.

% Then, it is interesting to look more precisely at the influence of each
% instrument. To achieve this goal, the lists of the removed instruments, which
% are known for a given scenario, need to be studied. For a core instrumented with
% $n$ instruments, there are  $C^{p}_{n}$ possible scenarios to remove $p$
% instruments. Looking at the list of instruments removed, a statistic about how
% many times an instrument appears within a set of lists of removed instruments
% can be built. We assume all the $C^{n}_{p}$ possible scenarios are considered.
% Thus each instrument will appear within the list of removed instruments the same
% number of times. Moreover, if all the instruments are equivalent, plotting those
% instrument occurences within an histogram will lead to a flat one.
% 
% Using the norm $\epsilon_i$ to select the scenarios of removed instruments, then
% we can have the same technique to analyse the influence of instruments within
% subsets of scenarios. As before, if the final histogram is flat, this means that all
% instruments have an equivalent contribution. On the contrary, if at least one
% peak appears it means some instruments are more present within this particular
% slice of $\epsilon_i$.

From that quality measurement, the aim is now to determine the effect of each
instrument on the data assimilation results. To archive this goal a detailed
study of the distribution of the $\epsilon_i$'s is done. Several networks of
instruments within the core are considered to obtain an overall understanding of
the influence of instruments.

\subsection{Standard PWR900 instruments repartition \label{sec:PWR900}}

The first case is a standard PWR900, with data assimilation done as described
previously. In this case, the measurement error is proportional to the measure
itself as described by Equation \ref{eqR}.

Here are studied the scenarios when only two instruments remain. In such case,
the histogram of the $\epsilon_i$'s is rather broad, as show in
Figure~\ref{fig:inf-fig1}. Large distributions (as the one of 2 instruments
lost) ensures a better separation of the different classes of instruments that
can be present within a given slice in the $\epsilon_i$'s values. In a very
narrow distribution, as the one where 48 instruments remains of
Figure~\ref{fig:inf-fig1}, confusion between the different classes associated to
a slice in $\epsilon_i$ value is more likely.

To quantify the results within a slice of $\epsilon_i$ values, first the whole
distribution is investigated  when no selection on $\epsilon_i$ is done. In this
case, the distribution of instruments occurrences as defined in Section
\ref{method} will be flat. The interesting point for further comparison is its
amplitude of $49$ occurrences. This value of amplitude represents $2450 = 1225
\times 2$ coupled references distributed over $50$ instruments. This amplitude
gives the reference amount to be compared with a slice of $\epsilon_i$ values. 

The subset of $10\%$ scenarios that give the highest values is selected to study
the influence of instruments among $\epsilon_i$ distribution. Assuming an equal
influence of all instruments in the $49$ cases, and considering the $10\%$
highest values of $\epsilon_i$, the occurence of each instrument should have
only a mean value of $4.9$ cases.

\begin{figure}[!ht] 
\begin{center} 
   \includegraphics[width=\textwidth]{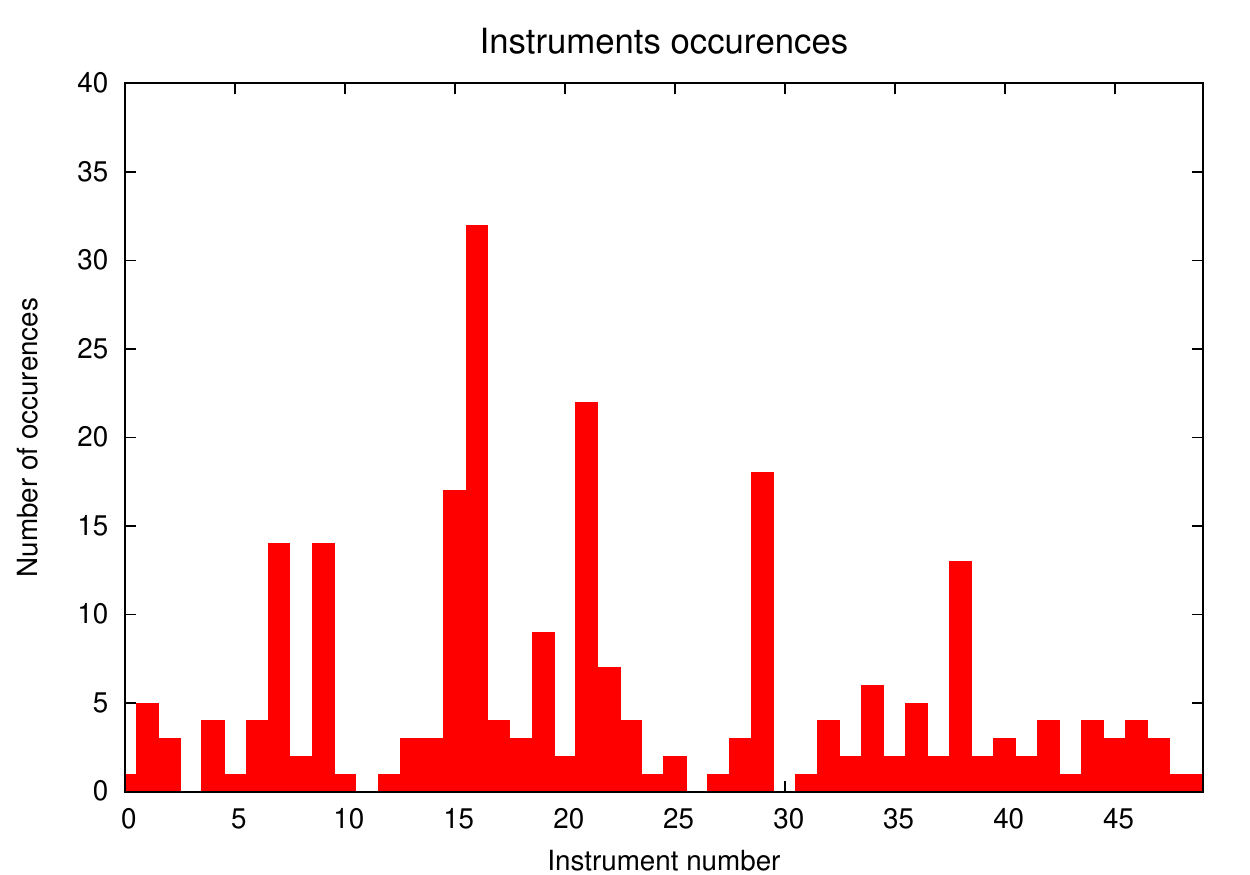} 

   \caption{Histogram of the occurrence of the instruments in the $10\%$
  scenarios that give the highest values of $\epsilon_i$. Each value in abscisse
  are corresponding to a reference number of one instrument. In ordinate is
  shown the occurrence of each instrument within remaining instruments list.   
  \label{fig:inf-fig2} }  

\end{center} 
\end{figure}

The histograms are built according to the method described in Section
\ref{method}. The result is plotted in Figure  \ref{fig:inf-fig2}.

\begin{figure}[!ht] 
\begin{center} 
  \includegraphics[width=\textwidth]{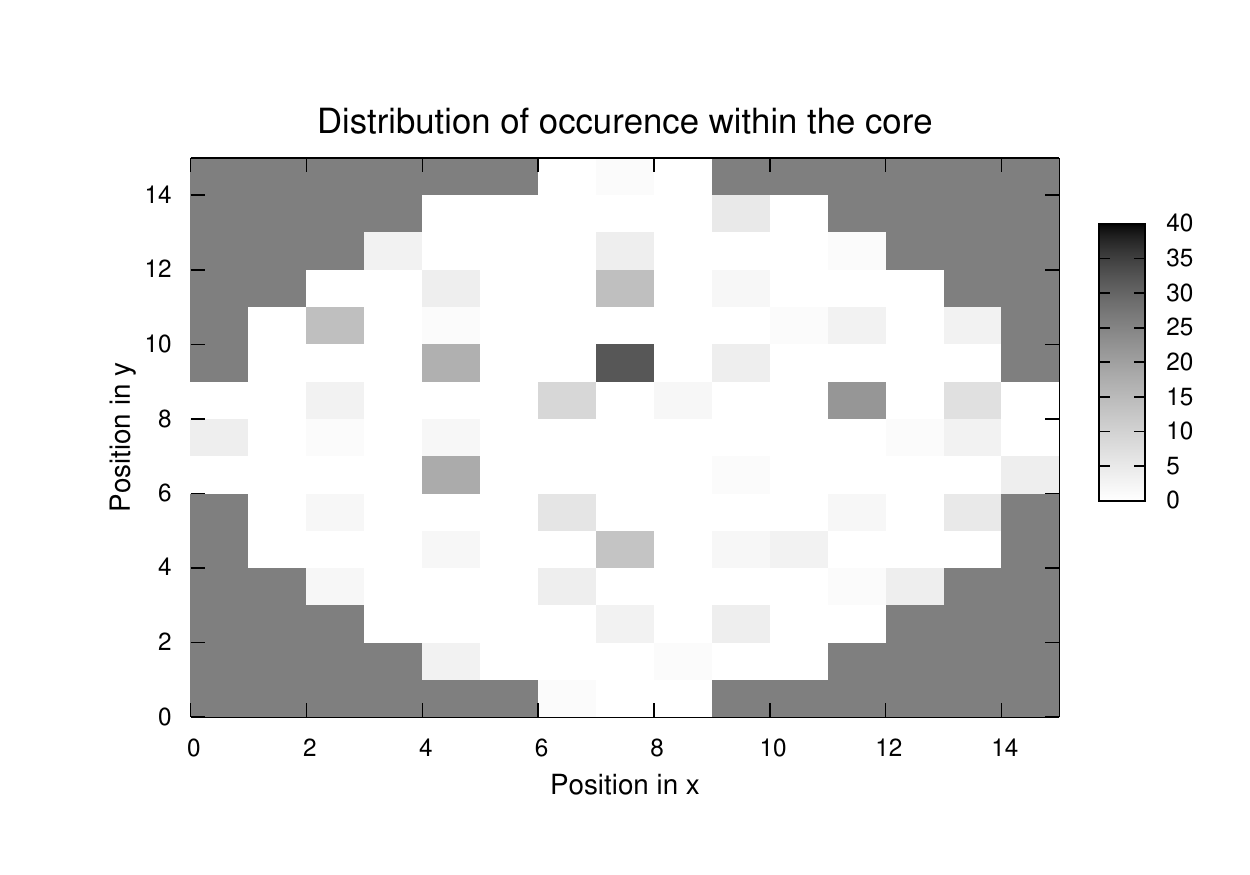}  

  \caption{Occurrence of the instruments as a function of their position with in
  core. The cases kept for doing this histogram are the ones within the $10\%$
  highest values of the $\epsilon_i$ distribution. \label{fig:inf-fig5} } 

\end{center} 
\end{figure} 

The distribution presented in Figure~\ref{fig:inf-fig2} is not flat, this
indicates that the instruments are not equivalent. Moreover the maximum value of
occurrence is $32$, which is significantly above the  $4.9$ mean value
prescribed in the case where all instruments are equivalent. The peaks on
Figure~\ref{fig:inf-fig2} are putting in light the instruments that are the
least useful for activity field reconstruction. The instruments   involved in
the low reconstruction quality slice are mainly localised in the centre of the
nuclear core as it can be noticed on Figure~\ref{fig:inf-fig5}.  One hypothesis
that can be advanced to explain those differences is the difference of the
assembly kind. Looking at the fuel loading pattern we noticed that those effect
cannot be explained mainly by the technical specification of the assembly. 

This central location of instruments, that are the most present in histogram
could be interpreted as a general law. However, the spatial repartition of MFC
as show in Figure~\ref{fig:figMFC} is complex. Thus, this does not allow to give
clear conclusions. However, it worth to notice, comparing  Figures
\ref{fig:figMFC} and \ref{fig:inf-fig5}, that instruments that are geometrically
close do not have necessary the same behaviour. This is the sign of an inner
complexity of instrument location effects.

Thus, other configurations need to be investigated to collect more information
on this effect. To go further two points need to be considered: 

\begin{itemize}

\item the hypothesis taken here is to have some measurement error that are
proportional to the absolute value of the measurement, as given by
Equation~\ref{eqR}. This implies that implicit error is higher at location when
activity is stronger.  

\item the instruments repartition is complex. This has the benefit to work in
cases that are close to real one. However in the same time it leads to a
difficult interpretation.

\end{itemize}

Thus more theoretical and idealistic configurations need to be considerated to
get rid of those both difficulties points and to answer better the question.

\subsection{Regular instruments repartition \label{sec:inf-greg}}

\begin{figure}[!ht] 
\begin{center} 
  \includegraphics[width=\textwidth]{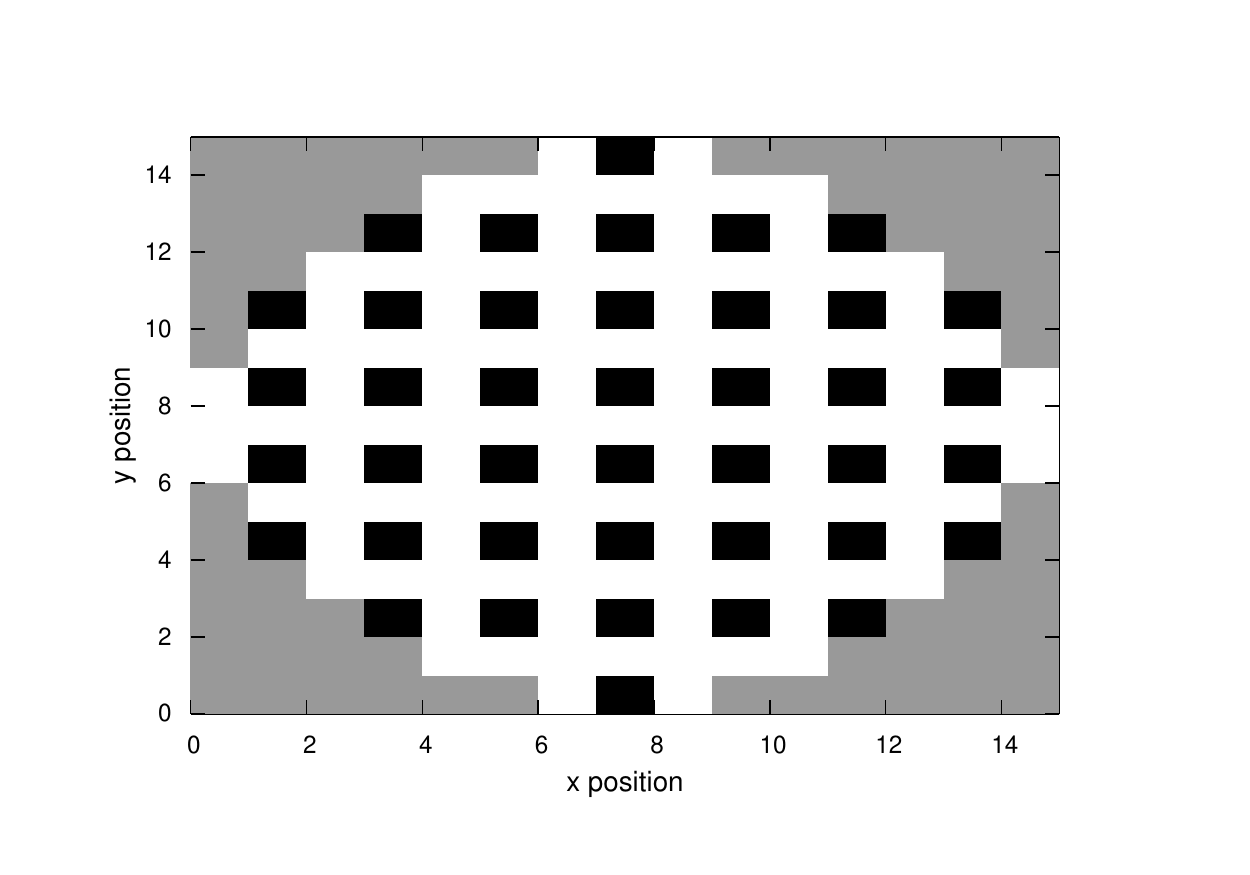} 

  \caption{The MFC instruments within the nuclear core are localised in
  assemblies in black within the horizontal slice of the core. The assemblies
  without instrument are marked in white and the reflector is in gray.
  \label{fig:inf-figReg}} 

\end{center} 
\end{figure}  

To avoid both difficulties presented in previous section, the following
instruments configuration is taken: 

\begin{itemize}

\item a repartition of the instruments within the core following a Cartesian
map. The location of the instruments within such a configuration is reported on
the Figure~\ref{fig:inf-figReg}. Within this configuration, only  $40$ MFC are
available. This number of instruments is a bit lower than the $50$ of the
standard PWR900 case presented in Figure~\ref{fig:figMFC}. It leads to an
overall difference in instruments density and quality of initial reconstruction.

\item Measurement errors are taken as constant, as described by
Equation~\ref{eqRflat}.

\end{itemize}

If only the hypothesis on geometry presented above is kept, the overall
conclusion of this section is not changed at all. This second hypothesis on
measurement error modelling, whatever is the instrument location choice, has
low effect on the results.

Using both above hypothesis for the calculation, the same study as the one done
for standard configuration, is done here. The analysis is done according to
method described in Section~\ref{method}. In the present case the value of
$\xaref$ is not the same as in the standard case described in
Section~\ref{sec:PWR900}. This new value of $\xaref$ is calculated assuming all
the $40$ instruments are localised according to Figure~\ref{fig:inf-figReg}.
Within such a configuration there are $C^{40}_{2}=780$ possible scenarios when
$2$ instruments remain,  so all configurations are evaluated. 

If all the cases are taken into account, an uniform distribution is obtained.
The amplitude of this distribution is $39$. This value represent $1560 = 780
\times 2$ coupled references distributed over $40$ instruments. This means that,
for a selection of $10\%$ of the scenarios the mean flat value is around $4.0$.
This value is a reference to compare with the values obtained in occurrence
distribution within a slice of value of  $\xa-\xaref$. 

\begin{figure}[!ht] 
\begin{center} 
  \includegraphics[width=\textwidth]{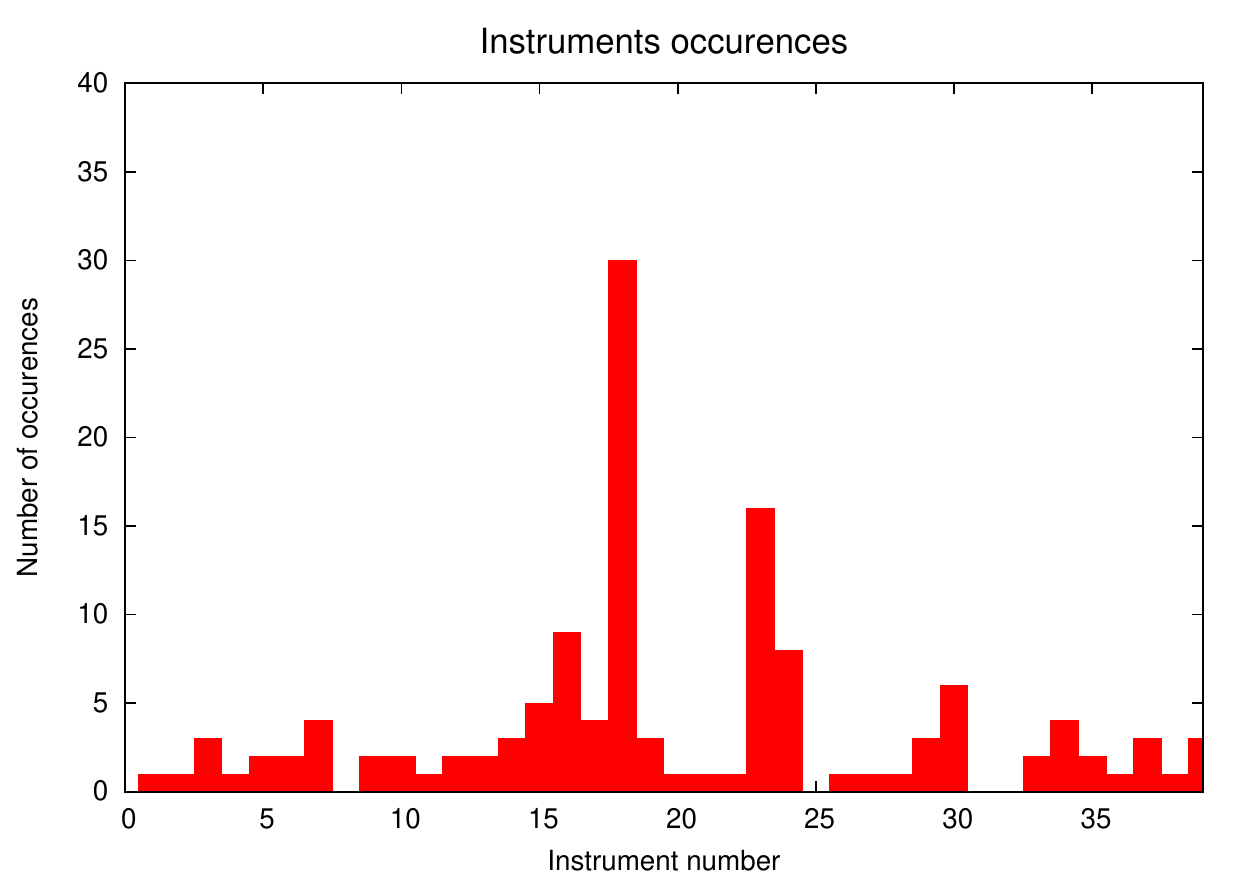} 

  \caption{Histogram of the occurrence of the instruments in the $10\%$
  scenarios that give the highest values of $\epsilon_i$. Each value in abscisse
  is corresponding to a reference number of one instrument. In ordinate is shown
  the occurrence of the instruments within remaining instruments list.   
  \label{fig:inf-fig11}}  

\end{center} 
\end{figure}

As in Section \ref{sec:PWR900},  the ensemble of scenarios that leads to the
$10\%$ highest $\xa-\xaref$  values are investigated. The occurrence of the
instruments for the scenarios giving the $10\%$ highest $\xa-\xaref$  values
are plotted in Figure~\ref{fig:inf-fig11}.

\begin{figure}[!ht] 
\begin{center} 
  \includegraphics[width=\textwidth]{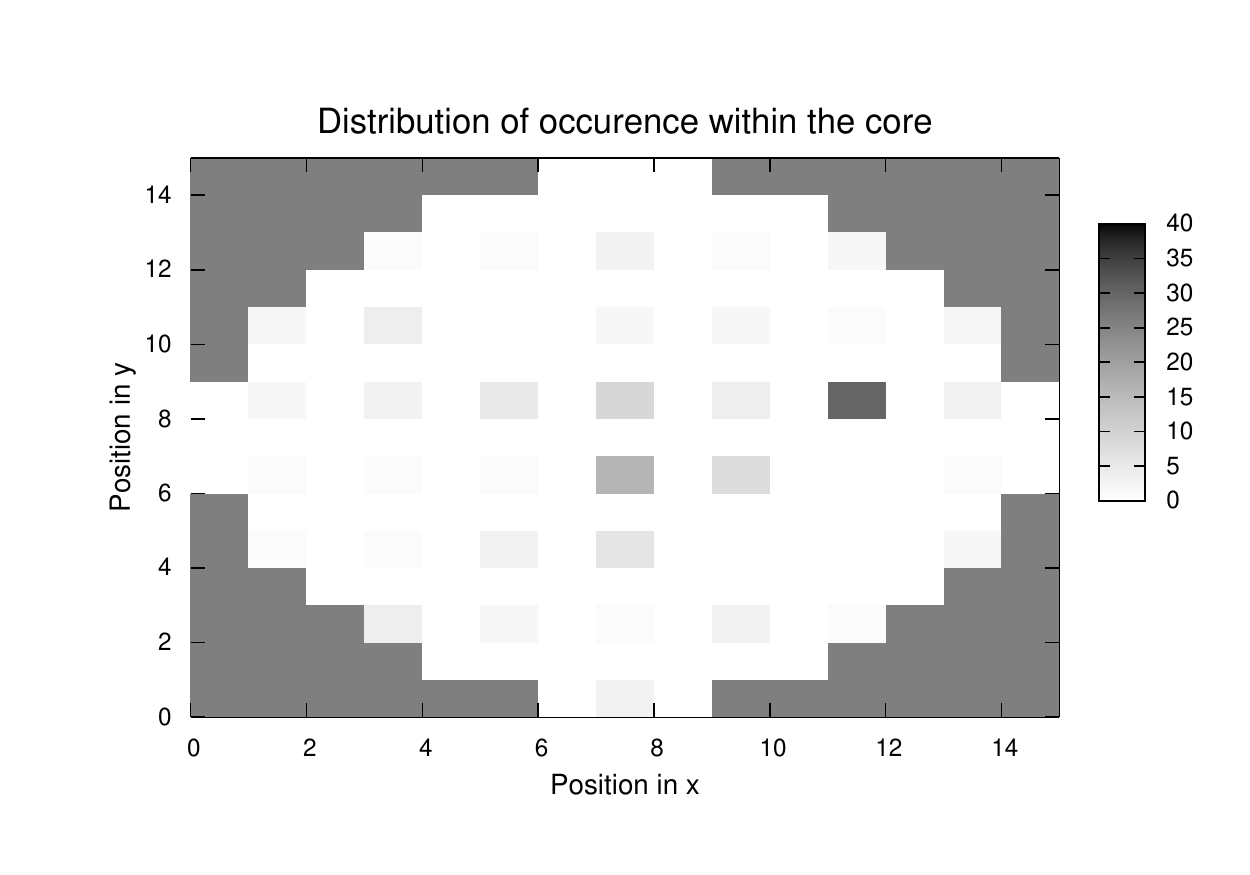}  

  \caption{Occurrence of the instruments as a function of their position within
  the core. The cases kept for doing this histogram are the ones within the
  $10\%$ highest values of the $\epsilon_i$ distribution.
  \label{fig:inf-fig14}} 

\end{center} 
\end{figure}

The peaks in Figure~\ref{fig:inf-fig11} within the distribution prove that the
instruments are not equivalent, even in the case of a regular repartition of
instruments according to a Cartesian grid. This confirms that all instruments do
not have the same effect, whatever is the chosen network. Another point is the
quasi symmetry of the distribution with respect to the mean of the chosen
numbering. The numbering of the instruments is done from right to left and from
top to bottom. Which means that the instruments numbered $1$ at the top left is
equivalent to the one numbered $40$ at the bottom right. This classification
lead to  equivalence between the instruments numbered from $1$ to $20$ with the
corresponding one, respectively, between $40$ and $21$. Thus symmetry imposed by
numbering appears in the histogram. Moreover, even if a fixed value of the
measurement error is imposed, this do not eliminate the effect of location on
the instruments. This location effect is then rather dominant with respect to
the error effect. To have a better view on a plan of those instruments, their
spatial locations within the core are represented in Figure~\ref{fig:inf-fig14}.

Figure~\ref{fig:inf-fig14} shows the instruments that have a low contribution
to the field reconstruction with data assimilation. Those instruments are
mainely localised in the centre of the core. In this case this interpretation
is easy as the instruments location are distributed regularly as shown in
Figure~\ref{fig:inf-figReg}.  This central location of the instruments confirms
what was seen in Figure  \ref{fig:inf-fig11} about the central instruments
locations. 

Beyond this global similitude, several differences are notable comparing
Figure~\ref{fig:inf-fig14} and Figure~\ref{fig:inf-fig5}. In
Figure~\ref{fig:inf-fig14}, the most represented instrument is located in the
top middle of the core centre. However, in Figure~\ref{fig:inf-fig11}, this
instrument is located on the right side of the core. This effect of location
changing, on the most represented instrument, can be attributed to the global
difference between  available instruments positions as measurement error
repartition and assembly technical specifications hypothesis can be excluded. No
effect of the modelling of the error respect to Equation \ref{eqR} or Equation
\ref{eqRflat} can be seen whatever is used. The repetition of the assembly is
exactly the same as in standard case and the assembly technical characteristics
are not correlated directly to the results.  

The conclusion from that point is that locations of the most influential
instruments are dominated by the instrumental network pattern chosen for the
core. Moreover, the precise location of the most represented instrument can be
put in light precisely using statistics on the occurrence of each instrument in
specific slice of $\epsilon_i$ norm values. In the study presented in Section
\ref{sec:PWR900} and \ref{sec:inf-greg}, both are assuming that two instruments
are added to a non instrumented core. This kind of hypothesis is rather strong
and do not allows to conclude in a more general case. 

From present step, it seems that the instruments located at the centre are the
most represented in the $10\%$ highest $\epsilon_i$ values. In this sense, they
lead to the worth reconstruction of the core, which seem very paradoxal. To
understand that point, it is necessary to keep in mind  that within a data
assimilation procedure, the measurement provided by an instrument have influence
within a rather large radius around the measure itself.  This comes from the
construction on the $\oB$ matrix that is presented in Equation \ref{eqB}.
Typically, this matrix is constructed with an influence length of few
assemblies. Thus, the improvement through data assimilation is driven by the few
measurements that can generated too large modification, that are not 
optimum on close assemblies. This does not lead to a overall quality improvement
of the final analysis. 

Such an effect means that it is necessary to have more instruments in the
starting network to make a better balance between information they provide. To
confirm this hypothesis, the amplitude of the highest peak in
Figure~\ref{fig:inf-fig11} give enough information. This plot correspond to  the
occurrence of a given instruments in the $10\%$ higher value of $\xa-\xaref$
norm. On Figure  \ref{fig:inf-fig11}  maximum peak value is $30$. This value is 
$23\%$  lower than the flat limit when no selection is done  at $39$. This mean
that in $23\%$  of the cases, there are a counter balancing of the
overconfidenced information given by one instrument. Thus, the influence of
other instruments in the network must be considered.

\subsection{Half instrumented core with regular repartition}

The aim of this section is to comfort conclusions on the influence of starting
instrumental network configuration obtained in Section \ref{sec:PWR900} and
\ref{sec:inf-greg}. To keep the advantage of the flat error and regular
distribution  new case is build on those basis. The assumption  that the core is
half instrumented will be done in this part. This means that one half will be
considered as fixed one. Then only instruments coming from the other half are
added. 

\begin{figure}[!ht] 
\begin{center} 
  \includegraphics[width=\textwidth]{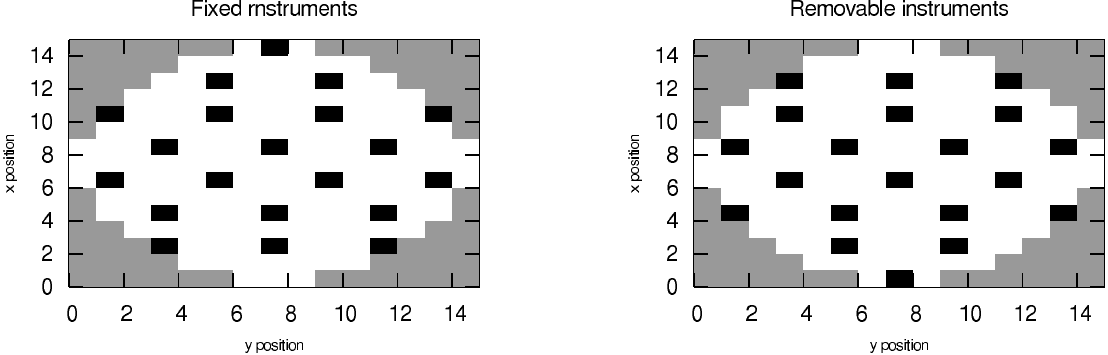} 
 
  \caption{The MFC instruments within the nuclear core are localised in
  assemblies in black with in the horizontal slice of the core. The assemblies
  without instrument are marked in white and the reflector is in gray. The left
  picture is for the instruments that are always kept. The right figure
  represent instruments that may be suppressed. \label{fig:inf-figSpc}} 

\end{center} 
\end{figure}

To keep the regularity of the distribution, instruments are separated in the
fixed or removable categories alternatively using as basis the regular
distribution presented in Section \ref{sec:inf-greg}. Thus, two categories  of
$20$ instruments each are obtained.  The location of the instruments for each
category is presented on Figure~\ref{fig:inf-figSpc}. This process to build
classes, even if it leads to some asymmetry, allows on overall to keep the
convenient regularity features of the initial repartition.

Thus, an addition of $2$ instruments from the removable set is done to the set
of $20$ fixed instruments. As only $20$ instruments remains the number of
possible scenarios is smaller than in Section \ref{sec:PWR900} and
\ref{sec:inf-greg} remain. In this case, there are only  $C^{2}_{20}$ cases
which means $190$ possibilities. As expected, the norm distribution of
$\xa-\xaref$ is a narrower that the previous case. This distribution, contrary
to the one presented for two instruments remaining in Figure~\ref{fig:inf-fig1},
is rather symmetric. Still the distribution is board enough to make a cut over
the $10\%$ higher values of the distribution. 

To get a clear view of the instruments, we will make occurrence histogram taking
into account only the two instruments that are added. However we will still
working in the framework of all instruments set. This new configuration will
be analysed according to the method used in Section \ref{sec:PWR900} and
\ref{sec:inf-greg}. 

\begin{figure}[!ht] 
   \begin{minipage}[t]{0.5\linewidth} 
   \centering
   \includegraphics[width=6cm]{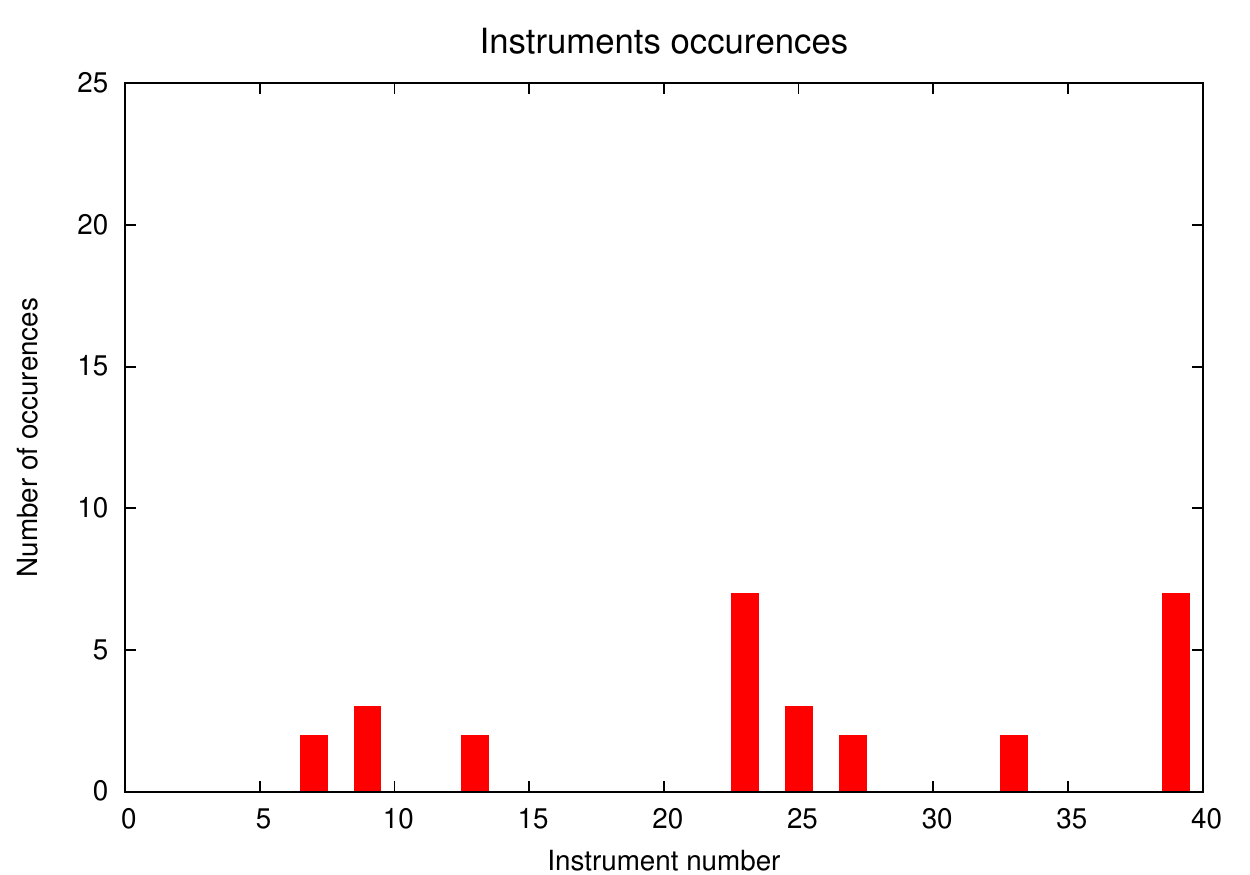} 
   
   \caption{Histogram of the occurrence of instruments for the $10\%$
   highest values of the $\epsilon_i$ distribution. Each value in abscisse are
   corresponding to a reference number of one instrument. In ordinate is shown
   the occurrence of the instruments within remaining instruments list.   
   \label{fig:inf-fig15} }  

  \end{minipage}
  \hspace{0.5cm}
  \begin{minipage}[t]{0.5\linewidth}
  \centering    
  \includegraphics[width=6cm]{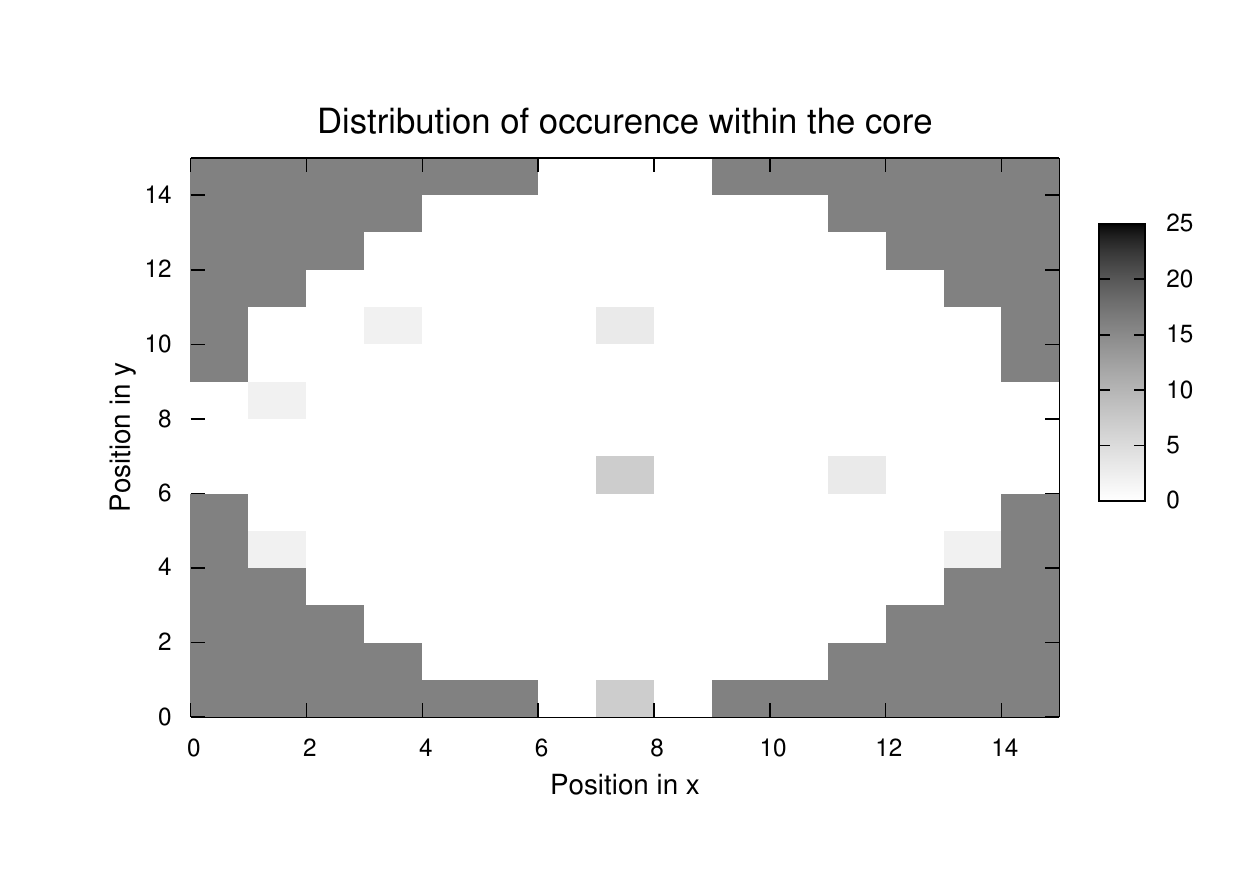}  
  \caption{Occurrence of the instruments as a function of their position with in
  core. The cases kept for doing this histogram are the ones within the $10\%$
  highest values of the $\epsilon_i$ distribution. \label{fig:inf-fig18}} 
 \end{minipage}
\end{figure}

In the present case only half of the instruments are considered. Thus the plot
of occurrence on all the studied scenarios is a sampling function as only the
locations with a removable instrument are considered. The frequency of this
sampling function is $1/2$  instrument. This characteristic frequency comes from
the regular alternative sorting of instruments  either  in removable set of
instruments either in fixed set of instruments. The amplitude of this sampling
function is $19$  which correspond to $190 \times 2 $ divided by the $20$
possibles locations. In this case, this is the difference respect of this
sampling function that will be studied. As previously, this is the $10\%$ higher
value of norm of $\xa-\xaref$  distribution that are investigated. The results
of the instruments occurrence histogram are plotted on
Figure~\ref{fig:inf-fig15}. For this $10\%$ subset of instruments, the
hypothesis of independence of instruments would leads to an histogram of
occurrence as a sampling function of frequency $1/2$ and mean amplitude of
$1.9$ 

In Figure  \ref{fig:inf-fig15} the pattern of occurrence is not any more a
sampling function with a constant amplitude but some peaks appear.  The
amplitude of the most important peak are around $7$ which is far beyond the mean
value of the regular sampling function that is $1.9$. Such a huge amplitude peak
signs one more time the non equivalence of all the instruments within the data
assimilation procedure. No more symmetry exists in the histogram comparing to
Figure~\ref{fig:inf-fig11} and this is even asymmetry that appears. This an
interesting change as the instruments selection  method should keep some how
this symmetry. This sign an influence of the existing instruments network on
data assimilation results. 

The precise spatial location of the instruments is reported in
Figure~\ref{fig:inf-fig18}. Comparing the location of the instruments in
Figure~\ref{fig:inf-fig18} respect to the case of regular repartition for the
same kind of slice in $\xa-\xaref$ presented in Figure~\ref{fig:inf-fig14}
several differences are obvious.

The most interesting point is that in the present case presented in
Figure~\ref{fig:inf-fig15} a peak arise at the very bottom of the core. Thus,
the MFC that got the lowest influence on reconstruction by data assimilation are
in this case not any more localised in the centre of the core.  This show one
more time a significant difference with results obtained especially in Section 
\ref{sec:inf-greg} that is dealing with a very close initial configuration.

Even if the same instrument locations are used originally, those results are
very different from the ones presented in Figure~\ref{fig:inf-fig14}. Thus we
see that both  the original present instrumentation as well as the location of
the instruments have an effect on the activity field reconstruction quality.

Examining together Figures \ref{fig:inf-fig5}, \ref{fig:inf-fig18} and
\ref{fig:inf-fig14}, we can notice that the instruments that are the least
important often seem to be localised in area where the density of instruments
present or potentially available is fairly large. 

\section{Conclusion}

One of the purpose of  this study was to make a study of the influence
of the instruments within a data assimilation procedure to reconstruct activity
in nuclear core. The quality of reconstruction is evaluated thought the norm of
$\xa-\xaref$ for all the possible combination of two instruments addition that
is  the best case in terms of statistic and analysis of the results. 

Focusing on the instruments leading to the $10\%$  highest value of $\epsilon_i$
when instruments are added to a non instrumented core of PWR900, it can be
noticed that they are not equivalent. It appears that within a distribution that
is supposed to be flat if all instrument play the same role, some instruments
have higher occurrence than other.

This effect can be understood focusing on a Cartesian regular repartition of
the instruments. Within this regular configuration, it was shown that the
instruments localised in the centre are the least important for activity field
reconstruction  with data assimilation. This proves that the instruments
location  is the most important factor that impact the quality reconstruction
with data assimilation.

To better understand  the effect of adding instruments to a system, we choose to
do do the same procedure starting from a core with a half instrumented regular
distribution. Through this study, it can be demonstrated that the locations in
the center of the least influential instruments is not a general rule. 

Within those studies, it is determined, empirically, that the influence of the
instruments is related to the density of the present or potentially available
instruments around this location.  

The determination of the worst instruments (respectively the best)  to add to an
measurement system, to improve data assimilation with it,  is depending on
several parameters. 

The starting instrumental configuration to which instruments need to be added
play a fundamental role. It was proven here that results are very  different
when instruments are added to a non instrumented system than to a partially
(half) instrumented one. Such a non equivalence respect to starting point proves
that building a complete instrumental system cannot be done iteratively. Such a
system will have limited efficiency as each step is dependant of the previous
and not of the global situation. 

This imply that, in order to build an optimal measurement network in a nuclear
core, it is necessary to  be able to take into account all the instruments
globally.  

Developing tools and diagnostic for a determination of such optimal  network is
then complex. This is especially true  when a lot of measurement are needed.
Moreover such a goal can only be achieved  through advanced mathematical study
and powerful computing usage.

% \bibliography{}
\bibliography{bibliographie}
\bibliographystyle{elsarticle-num}

\end{document}